\documentstyle[prb,preprint,aps]{revtex}

\begin{document}
\draft
\title{Soft modes in two- and eight-direction order-disorder 
ferroelectrics }
\author{Zhi-Rong Liu\footnote{Electronic address: zrliu@phys.tsinghua.edu.cn} 
and Wenhui Duan}
\address{Department of Physics, Tsinghua University,
Beijing 100084, People's Republic of China}
\author{Bing-Lin Gu} 
\address{Center for Advanced Study, and Department of Physics, 
Tsinghua University, 
Beijing 100084, People's Republic of China}
\author{Xiao-Wen Zhang}
\address{State Key Laboratory of New Ceramics and Fine Processing,
Department of Materials Science and Engineering, 
Tsinghua University, Beijing 100084, People's Republic of China}
\maketitle

\begin{abstract}
The soft modes in the two- and eight-direction order-disorder ferroelectrics 
are calculated under the mean-field approximation. We find that 
the conventional 
method of the pseudospin model to calculate the soft-mode frequency 
is {\it incorrect}, and present a valid modified method. 
It is demonstrated that the conventional method does not show the soft-mode frequency 
going to zero at a critical temperature in the presence of random internal 
fields, while the frequency calculated by our modified method goes to zero in 
random fields at a critical temperature. In the 
eight-direction ferroelectrics, the soft-mode frequency decreases to zero at 
the first-order phase transition temperature though the symmetry has 
been destroyed at high 
temperatures under an external field. The promotion effect of random 
fields on the phase transition is testified by the calculation results of the 
soft modes in the paraelectric phase.

\end{abstract}

%

\vspace{2mm}


\section{Introduction}

A pseudospin model is usually used to interpret the phase-transition 
phenomena of order-disorder ferroelectrics such as KH$_2$PO$_4$ in terms of the 
microscopic atomic interactions.\cite{1} The model describes the ferroelectric phenomena 
as the motion of active ions in 
the double-well type potential field. The dynamic properties are 
generally considered by calculating the frequency of the ferroelectric soft mode. It 
yields satisfactory results for the second-order ferroelectric phase 
transition: when the temperature increases, the frequency of the soft mode 
decreases below the critical temperature $T_{\rm c}$ and increases above $T_{\rm c}$. 
At the critical temperature $T_{\rm c}$, the frequency is equal to zero, i.e., 
the ion vibration is frozen. In the previous works,\cite{1,1a} the conventional equation 
to determine the frequency of the soft mode is expressed as 
\begin{equation}
\frac{d\langle {\bf S}_i \rangle _t}{dt}=-i\langle [{\bf S}_i,H] \rangle _t,
\end{equation}
where ${\bf S}_i$ is the pseudospin of the $i$-th ion, $H$ is the Hamiltonian 
of the system, and $t$ is 
the time. $\langle \ldots \rangle _t$ denotes the temporal thermal averaging. 
The equation (1) is obtained by making thermal averaging on the quantum mechanics equation 
\begin{equation}
\frac{d{\bf S}_i}{dt}=-i[{\bf S}_i,H].
\end{equation}
To obtain 
Eq. (1) from Eq. (2), an assumption implied here is 
\begin{equation}
d\langle {\bf S}_i \rangle _t=\langle d{\bf S}_i \rangle _t,
\end{equation}
i.e.,
\begin{equation}
d\langle {\bf S}_i \rangle _t=\sum\limits_{\alpha}\rho_{\alpha}d{\bf S}_{i\alpha},
\end{equation}
where ${\bf S}_{i\alpha}$ is the spin of the $\alpha$-th eigen state, and 
$\rho_\alpha$ is the 
statistic weight. However, starting from the original definition of 
$\langle {\bf S}_i \rangle _t$, i.e., 
\begin{equation}
\langle {\bf S}_i \rangle _t=\sum\limits_{\alpha}\rho_{\alpha}{\bf S}_{i\alpha},
\end{equation}
the correct equation should be
\begin{equation}
d\langle {\bf S}_i \rangle _t=\sum\limits_{\alpha}
    [\rho_{\alpha}d{\bf S}_{i\alpha}+{\bf S}_{i\alpha}d\rho_{\alpha}].
\end{equation}
Therefore, one purpose of this article is to investigate the soft mode 
of the order-disorder ferroelectrics by using Eq. (6) and compare with the 
results of the conventional method.

Rather different from the normal ferroelectrics, relaxor ferroelectrics (relaxors) 
experience no macroscopic phase transition at zero external field.\cite{2,3} 
However, a strong 
external electric field can induce a first-order ferroelectric phase 
transition in the relaxors such as Pb(Mg$_{1/3}$Nb$_{2/3}$)O$_3$ (PMN).
\cite{3,4,5} The phase transition and other 
special characteristics (such as the strong frequency dispersion and the 
history-dependent polarization behaviors) of relaxors are widely believed to be caused 
by the random interactions and the random electric fields in the system.\cite{6,7,8,9} 
Based on the experimental observation\cite{10} that the polar active ions in PMN 
shift along eight \{111\}-equivalent directions, 
an eight-direction order-disorder ferroelectric  model\cite{11} was presented to 
explain the induced first-order phase transition. 
The dynamic properties, however, are not calculated in that work.\cite{11} 
So, another purpose of this article is to examine the dynamic 
properties of the phase transition in the 
eight-direction order-disorder ferroelectric system.

\section{Methods}

In the eigen state space of the operator $S^z$, the Hamiltonian of the 
two-direction pseudospin model
can be equivalently expressed as 
\begin{equation}
H=\left[
\begin{array}{cc}
-Ep_0 &  {\frac{\Omega}{2}} \\
\frac{\Omega}{2} & Ep_0 
\end{array} \right],
\end{equation}
where $E$ is the electric field strength, and $p_0$ is the magnitude of 
the dipole moment when an ion locates in a certain potential 
well. $\Omega$ is the tunneling frequency between the two potential wells, 
whose magnitude 
is equal to the energy difference between the asymmetric 
and the symmetric eigen states at zero field. 
The dipole moment $p$ is associated with $S^z$ through the relation 
\begin{equation}
p=2p_0S^z.
\end{equation}
Under the mean field approximation, the interactions upon a certain 
ion may be represented by an equivalent field, i.e., 
\begin{equation}
Ep_0=J\frac{\langle p \rangle}{p_0}+E_{\rm ext}p_0+E_{\rm rand}p_0,
\end{equation}
where $\langle p \rangle$ is the thermal average value of the dipole 
moment, and $J$ is the ferroelectric 
coupling energy between polar ions. $E_{\rm ext}$ is the 
applied external electric field, and $E_{\rm rand}$ is the internal random fields 
in the system. 
$E_{\rm rand}$ is necessary to explaining the phase transitions in incipient 
ferroelectrics such as KTaO$_3$:Li,Nb,Na; SrTiO3:Ca or PbTe:Ge with dipole 
impurities (active ions with random sites and orientations).\cite{9} 
It comes from the direct interactions of electric dipoles, fields created 
by point charge defects and aforementioned defects, etc. Random fields 
are also important to interpreting the special properties of ralaxors such as 
PMN.\cite{6,7,8} The pinning effect in PMN was believed to be induced by 
the random fields of impurities in the system.\cite{8} In the process of the 
thermally activated flips of the local spontaneous polarization, the random 
interactions between polarization microregions are essential in producing 
the ralaxor characteristics such as the diffused phase transition and the 
frequency dispersion.\cite{7} It is also revealed that the charged chemical 
defects and nano-domain textures originating from 
atomic ordering also give important contributions to the random fields.\cite{12} 
The real occurrence of long-range ordered phase depends on the competition of 
the constant-sign ferroelectric coupling and the alternating-sign random fields. 
The amplitude of random fields is mainly determined by the impurity and the microstructure 
imhomogeneity in the system. If there are more charge impurities or the 
microstructure is more inhomogeneous, the random fields are stronger. 
Random fields coming from different sources have different 
distribution,\cite{7,9} but have similar influences. 
In this article, $E_{\rm rand}$ is assumed 
to have a Gaussian distribution with a width $\sigma_{\rm e}$ as  
\begin{equation}
\rho(E_{\rm rand})=\frac{1}{\sqrt{2\pi}\sigma_{\rm e}}
\exp\left[\frac{-E_{\rm rand}^2}{2\sigma_{\rm e}^2}\right].
\end{equation} 
The static properties of the system can be easily 
determined from Eqs. (7-10), which have been discussed 
extensively previously.\cite{1,9}

To investigate the dynamic properties of the system, the fluctuation 
of the electric field 
\begin{equation}
dE(t)=dE\cdot \exp (i\omega t)
\end{equation}
is considered. 
The eigen energy and the eigen polarization (dipole moment) of the Hamiltonian 
in Eq. (7) are respectively expressed as 
\begin{equation}
E_{\pm}=\pm \sqrt{(Ep_0)^2+(\Omega/2)^2}, 
\end{equation}
and
\begin{equation} p_{\pm}=\frac{Ep_0}{E_{\pm}}p_0.
\end{equation}
The variation of $p_{\pm}$ induced 
by $dE(t)$ can be calculated in a linear perturbation method of the quantum
mechanics, which yields
\begin{equation}
dp_{\pm}(t)=\frac{\Omega^2}{(Ep_0)^2} \cdot
             \frac{E_{\pm}p_0dE(t)}{4E_{\pm}^2-(\hbar\omega)^2}\cdot p_0.
\end{equation} 
According to the conventional method [Eq. (4)], $d\langle p \rangle _t$ reads as 
\begin{equation}
d\langle p \rangle _t=\int dp_+(t)\tanh\left(\frac{E_+}{k_{\rm B}T}\right)\cdot
                        \rho(E_{\rm rand})dE_{\rm rand} , 
\end{equation}
while in the modified method, $d\langle p \rangle _t$ reads as 
\begin{equation}
d\langle p \rangle _t=\int \left\{ dp_+(t)\tanh\left(\frac{E_+}{k_{\rm B}T}\right)+
            p_+\left[1-\tanh^2\left(\frac{E_+}{k_{\rm B}T}\right)\right]
              \frac{Ep_0^2dE(t)}{k_{\rm B}TE_+}\right\}
                        \rho(E_{\rm rand})dE_{\rm rand} . 
\end{equation}
Thus the frequency of 
the soft mode, $\omega$, can be determined from the self-consistent 
equation
\begin{equation}
p_0dE(t)=J\frac{d\langle p \rangle _t}{p_0}.
\end{equation}
The dynamic properties of the pseudospin model have been investigated in 
details by using the conventional method when there is no random field.\cite{1,1a}

For the eight-direction order-disorder ferroelectric model, 
there are eight potential wells along \{111\}-equivalent directions, 
and the Hamiltonian is assumed to be the 
direct sum of four groups of double-well matrix\cite{11}, i.e., 
\begin{equation}
H=\left[
\begin{array}{cccccccc}
-Ep_0 & \frac{\Omega}{2} & 0 & 0 & 0  & 0 & 0 & 0\\
\frac{\Omega}{2} & Ep_0 & 0 & 0 & 0  & 0 & 0 & 0\\
0 & 0 & -\frac{1}{3}Ep_0 & \frac{\Omega}{2} & 0 & 0 & 0  & 0 \\
0 & 0 & \frac{\Omega}{2} & \frac{1}{3}Ep_0 & 0 & 0 & 0  & 0 \\
0 & 0 & 0 & 0 & -\frac{1}{3}Ep_0 & \frac{\Omega}{2} & 0 & 0 \\
0 & 0 & 0 & 0 & \frac{\Omega}{2} & \frac{1}{3}Ep_0 & 0 & 0 \\
0 & 0 & 0 & 0 & 0 & 0 & -\frac{1}{3}Ep_0 & \frac{\Omega}{2} \\
0 & 0 & 0 & 0 & 0 & 0 & \frac{\Omega}{2} & \frac{1}{3}Ep_0  
\end{array} \right],
\end{equation}
where the factor 1/3 is the cosine of the angle between the dipole 
moment and the external field when they are not parallel to each 
other. Other equations similar to the two-direction cases can also be obtained, 
which are not listed here in order to keep the article concise.

\section{Results and Discussions}
\subsection{Two-direction pseudospin model}

In order to demonstrate the different effects of the conventional and the 
modified methods, 
the soft-mode frequency of the two-direction pseudospin model corresponding to 
$\Omega=J$ is shown in 
Fig. 1 when there are no external field and random fields. The 
dashed line represents the calculation results by using the conventional method, 
and the solid line represents those calculated by using
the modified method. We can see that both the conventional method
and the modified method yield the same frequencies of 
the paraelectric phases ($T>T_{\rm c}$). 
The reason is that the 
polarization and $z$-spin of the eigen states in the 
paraelectric phase are zero [$p_{\pm}=0$ while $E=0$. See Eq. (13)], 
so the Eq. (4) is equivalent to Eq. (6) in 
this case. For the ferroelectric phase, $(\hbar\omega)^2$ 
in the conventional method is proportional to the polarization of the system;\cite{1} 
$(\hbar\omega)^2$ in the modified method is smaller than that 
in the conventional method. 
The feature can be understood as follows: under the same drive 
$dE(t)$, the ``displacement" $d\langle {\bf S}_i \rangle _t$ calculated 
by Eq. (4) is smaller than that calculated by Eq. (6), i.e., the ``elastic" 
coefficient of the conventional method is larger than that of the modified 
method, so the 
vibration frequency in the conventional method is 
higher than that in the modified method. It 
should be noted that the frequency decreases to zero at the 
critical temperature, which is just a unique characteristic of the soft 
mode theory.

Figure 2 shows the polarization and 
the soft-mode frequency of the 
system under a weak external field $E_{\rm ext}=0.005 J/p_0$ when 
$\Omega=J$. It can be seen that the polarization increases continuously when 
the temperature decreases.  
The soft-mode frequency does not decrease 
to the zero value but reach a nonzero minimum. The minimal value of the 
frequency in the modified method is closer to zero than that in the 
conventional method. The frequency 
calculated in the modified method reaches its minimum at $T_{\rm m}'=0.95 J/k_{\rm B}$, 
which is a little larger than the temperature at which the polarization 
changes most quickly, $T_{\rm m}=0.91 J/k_{\rm B}$.

The random fields existing in the system have great influence 
on the properties of disordered ferroelectrics.\cite{9} They can make 
the critical temperature decrease, and even inhibit the phase transition. 
The polarization is calculated for the two-direction model with distribution 
width of random fields $\sigma_{\rm e}=0$ and 0.3 $J/p_0$, respectively, 
and the results are presented in 
Fig. 3(a). It is 
demonstrated that the critical temperature $T_c$ decreases from $0.911$ 
to 0.802 $J/k_{\rm B}$ when the random field width increases from 0 to 0.3 $J/p_0$. 
The figure 3(b) shows the calculated results of the soft-mode 
frequency when $\sigma_{\rm e}=0.3 J/p_0$ and $\Omega=J$. We observe
an extraordinary difference between the results calculated by the two 
methods: the frequency in the conventional method keeps nonzero while the frequency 
in the modified method decreases to zero at the critical temperature. 
From the curve of the polarization in Fig. 3(a) we 
know that the phase transition is of the second-order type and 
the symmetry of the system changes at the critical temperature. 
A zero frequency at the critical temperature is required by the soft 
mode theory. Thus, Eqs. (1) and (4) used in the past to calculate 
the frequency are incorrect, while Eq. (6) produces the 
physically reasonable results.

\subsection{Eight-direction order-disorder ferroelectric model}

Under the two-body couplings, there is only the second-order 
phase transition in the two-direction ferroelectric system, while both 
the second-order and the first-order phase transitions can exist in 
the eight-direction ferroelectrics.\cite{1,11} The polarization and the 
soft-mode frequency in both cooling and heating processes are 
shown in Fig. 4 for the eight-direction model when $\Omega=0.63J$. 
Fig. 4(a) clearly demonstrates the jump of the polarization and 
the difference of the phase transition temperature between the cooling 
and heating processes, which indicates 
that the phase transition is of the first-order type. In a cooling 
process, the soft mode frequency decreases to zero at the phase 
transition temperature (over-cooled temperature) and suddenly jumps to the 
nonzero frequency value of the ferroelectric phase. In a heating process, 
the cases are similar except that the phase transition occurs at a 
higher temperature (overheated temperature).

For the relaxor ferroelectrics such as PMN, an 
external field with proper strength will induce a first-order phase 
transition.\cite{3,11} The second-order phase transition occurs in stronger 
field ranges, and the phase transition vanishes when the field increases further. 
The curves of the soft-mode frequency at different external fields are 
depicted in Fig. 5 when $\Omega=J$. At the external fields $E=0.075$ and 
0.095 $J/p_0$, the phase transition is of the first-order type (see Ref. 12). 
There is a jump of frequency at the phase-transition temperature, and 
a down-pulse reaching the zero value 
exists above the phase transition temperature. 
When an external field is applied, the symmetry of the system has already been 
lowered at high temperatures, and the high and low temperature phases can not be 
distinguished from the symmetry changing. However, the zero soft-mode frequency 
at the critical temperature clearly indicates the existence of 
the phase transition. For 
$E=0.115J/p_0$, the phase transition is of the second-order type, 
and a down-peak of the frequency is easy to be identified. 
It is not a ``real'' phase transition in the sense that the soft-mode frequency 
does not decrease to zero. The phase transition can only be approximately defined from 
the rapid change of the polarization or from the minimum of the soft-mode frequency. 
For a strong field 
$E=0.15J/p_0$, the soft-mode frequency varies rather smoothly, 
and the softness of frequency is indefinite. 
In this case, it is difficult to identify any phase transition; in other words, 
the transition does not exist.

Rather different from the common sense, the random fields with proper distribution 
width will promote the spontaneous phase transition in the eight-direction 
ferroelectric system.\cite{11} In Fig. 6, the soft-mode frequencies of 
the paraelectric phase at different temperatures when $\Omega=J$
are given as functions of the width of random fields. 
With increasing width of the random fields, the frequency 
decreases first, and then increases. The minimal frequency is 
reached at a nonzero width. The promotion effect of 
random fields on the phase transition is demonstrated most clearly 
in the case of $T=0.085J/k_{\rm B}$: the random fields with 
distribution width $0.276J/p_0<\sigma_{\rm e}<0.436J/p_0$ induce the spontaneous 
appearance of the ferroelectric phase.

The original motivation of the eight-direction order-disorder ferroelectric model 
is to predict the phase transition in relaxors, but not to explore the dielectric mechanism.\cite{11} 
It is known that relaxors have a rather broad and smooth spectrum of 
the relaxation times, and the relaxation time shows a progressive slowing 
down upon cooling, which results in a transition into a glassy state.\cite{2,6,7} 
A distribution of relaxation time instead of a single frequency  
of the soft mode should be introduced for fully 
describing the dielectric properties of relaxors such as the frequency dispersion. 
The Monte Carlo simulation on the two-direction pseudospin model has been done to 
explain the dielectric behaviors of relaxors.\cite{7,13} Thus further works 
of simulation on the eight-direction model may be helpful for understanding 
the characteristics of relaxors.

\section{Summaries}

In summary, the soft modes in the two-direction and eight-direction 
order-disorder ferroelectric systems are syudied in this 
article. We indicate that the conventional 
method yields incorrect soft-mode 
frequency of the pseudospin model, and present the valid modified 
equation. The soft-mode frequency of the ferroelectric phase 
calculated by the modified method is smaller than the conventional 
values, while 
the critical temperature does not vary. When there exists random electric 
fields, the frequency calculated by the conventional method does not decrease to 
zero at the critical temperature, 
which is physically unreasonable. In contrast, the modified method yields 
the reasonable results. In 
the eight-direction ferroelectrics, the soft mode of paraelectric 
phase decreases 
to zero at the over-cooled temperature. The calculation of the soft modes 
under external electric fields shows that the frequency decreases to 
zero at the first-order phase transition temperature although the symmetry 
of the system has been destroyed at high temperatures. 
We find that 
the soft-mode frequency decreases first and then increases with 
increasing distribution width of the random fields, which 
testifies the promotion effect of random fields on the spontaneous 
appearance of the ferroelectric phase.

\section*{Acknowledgment}

This work was supported by the Chinese National Science Foundation 
(Grant NO. 59995520) and State Key Program of Basic Research 
Development (Grant No. G2000067108).

\begin{figure}[tbp]
\caption{Soft-mode frequency of the two-direction pseudospin model 
as function of temperature when $\Omega=J$. The dashed and the solid lines 
represent the results in the conventional and the modified methods, 
respectively. The temperatures 
is measured in unit of $J/k_{\rm B}$, and $\hbar\omega$ in unit of $J$. }
\end{figure}

\begin{figure}[tbp]
\caption{Temperature dependence of the polarization and the soft-mode 
frequency of the two-direction ferroelectrics at an external field 
$E_{\rm ext}=0.005J/p_0$ when $\Omega=J$. The polarization $\langle p \rangle$ 
is measured in unit of $p_0$. The dashed and the solid lines 
represent the results in the conventional and the modified methods, 
respectively. }
\end{figure}

\begin{figure}[tbp]
\caption{(a) The polarization of the two-direction ferroelectrics 
when $\sigma_{\rm e}=0$ (dot-dashed) and 0.3 $J/p_0$ (solid), respectively. 
(b) Soft-mode frequency of the two-direction ferroelectrics 
as function of temperature when $\Omega=J$ and $\sigma_{\rm e}=0.3J/p_0$, 
where the dashed and the solid lines 
represent the results in the conventional and the modified methods, 
respectively. }
\end{figure}

\begin{figure}[tbp]
\caption{Temperature dependence of (a) the polarization and (b) the soft-mode 
frequency of the eight-direction ferroelectris when $\Omega=0.63J$. 
The solid and dashed lines 
represent the results in a cooling process and a heating process, respectively. }
\end{figure}

\begin{figure}[tbp]
\caption{Soft-mode frequency of the eight-direction ferroelectrics 
as function of temperature at a cooling process when $\Omega=J$. 
Curves 1-4 correspond to $E_{\rm ext}$=0.075, 0.095, 0.115, and 0.15 $J$, 
respectively. }
\end{figure}

\begin{figure}[tbp]
\caption{Soft-mode frequency of the eight-direction ferroelectrics 
as function of random field distribution width when $\Omega=J$. 
The solid, dashed, and dot-dashed lines correspond $T$=0.1, 0.09, and 0.08 $J/k_{\rm B}$, 
respectively. }
\end{figure}

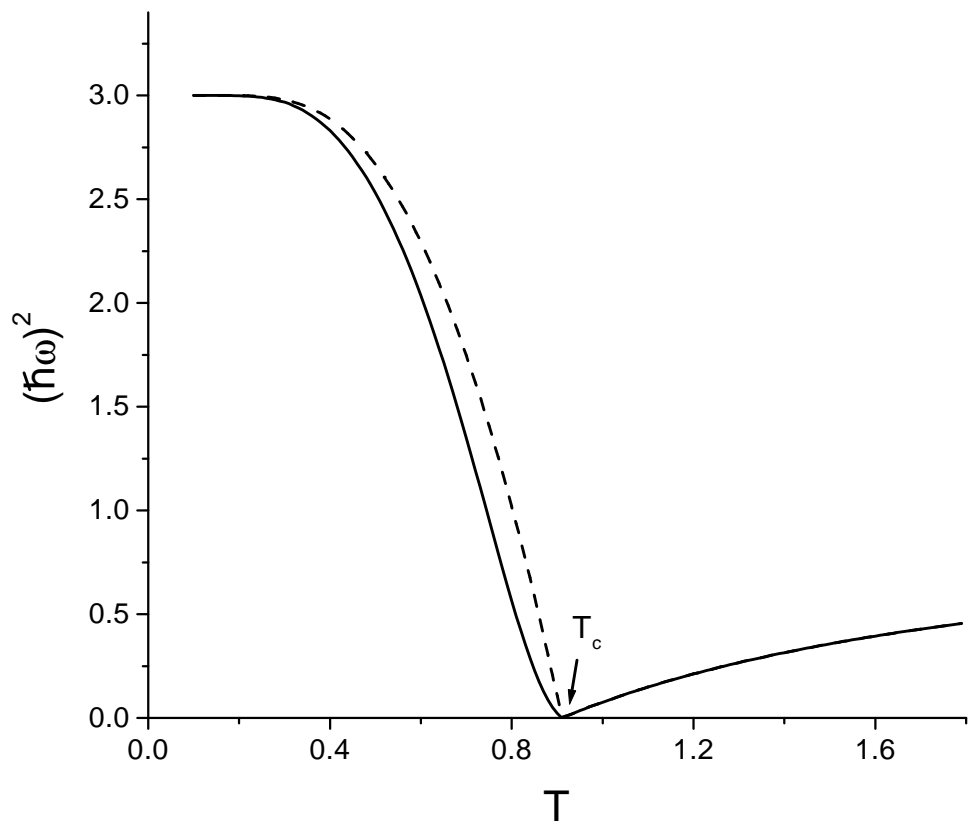
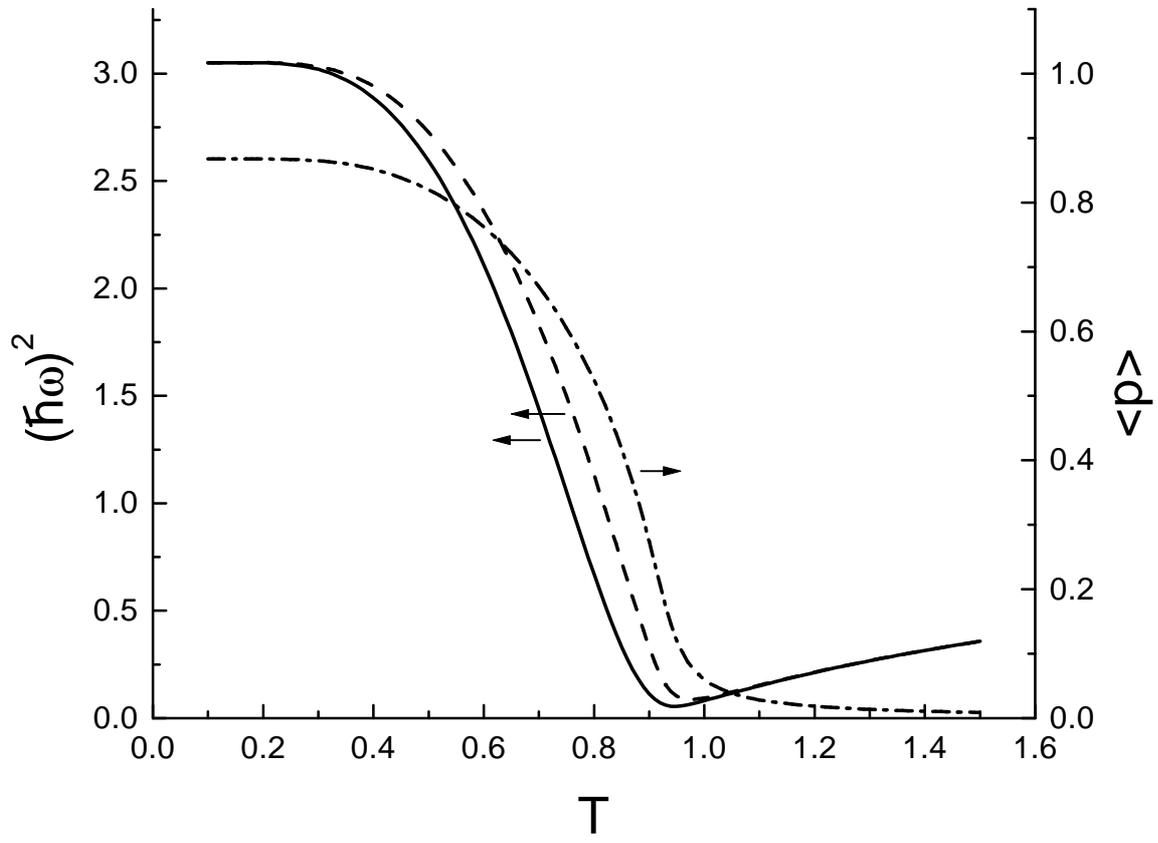
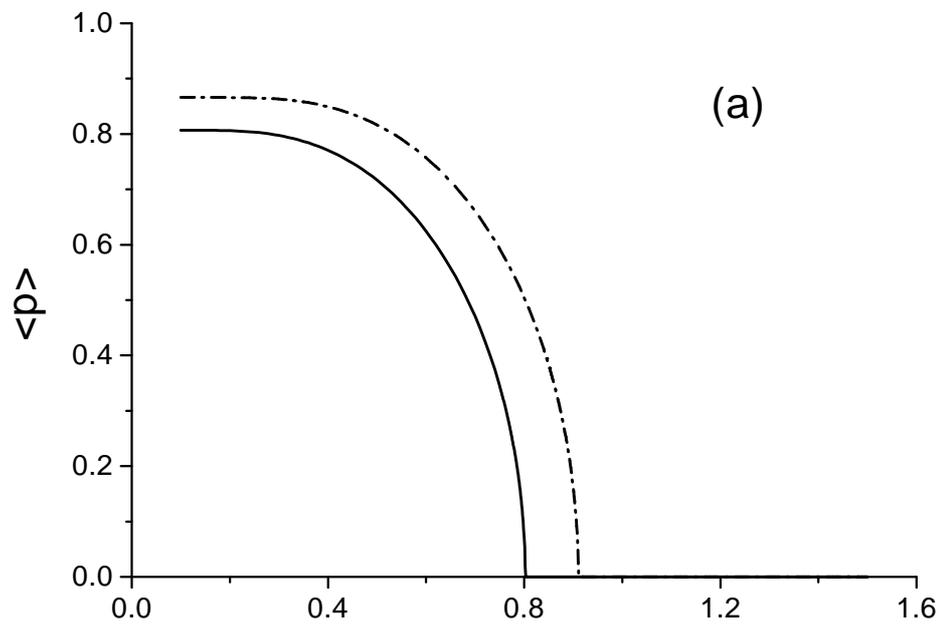
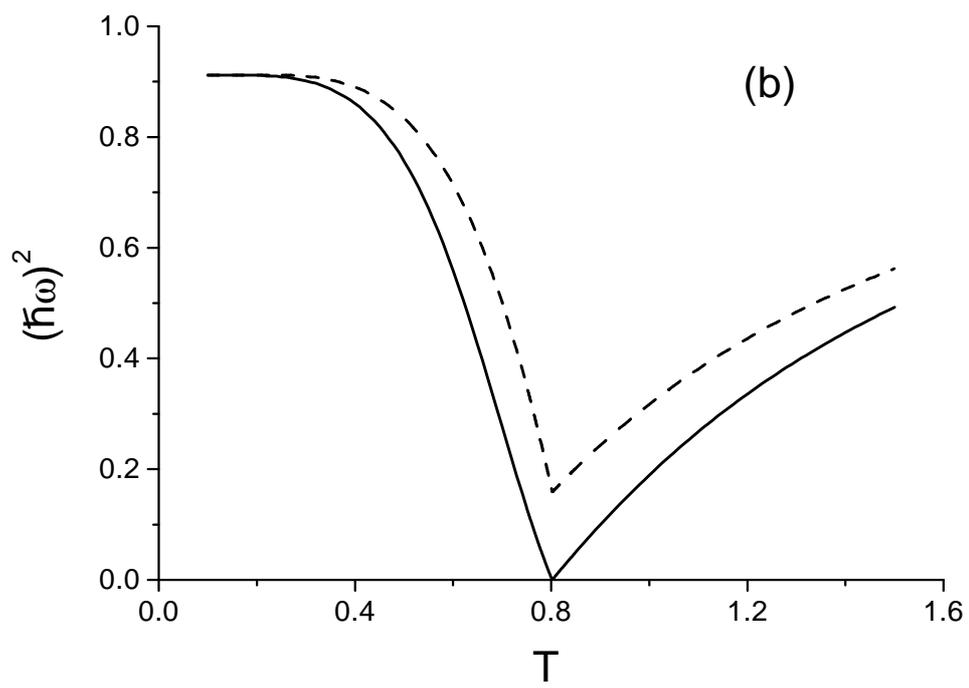
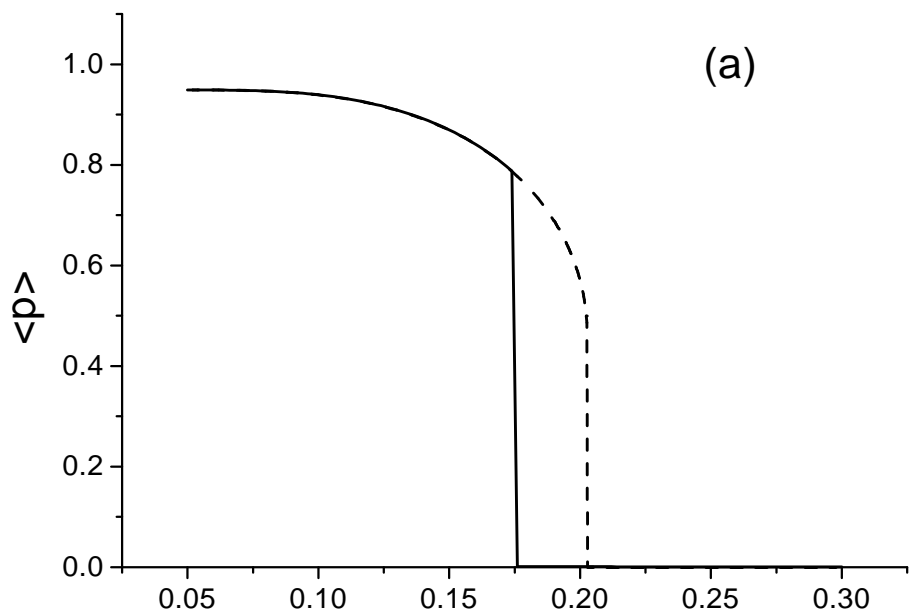
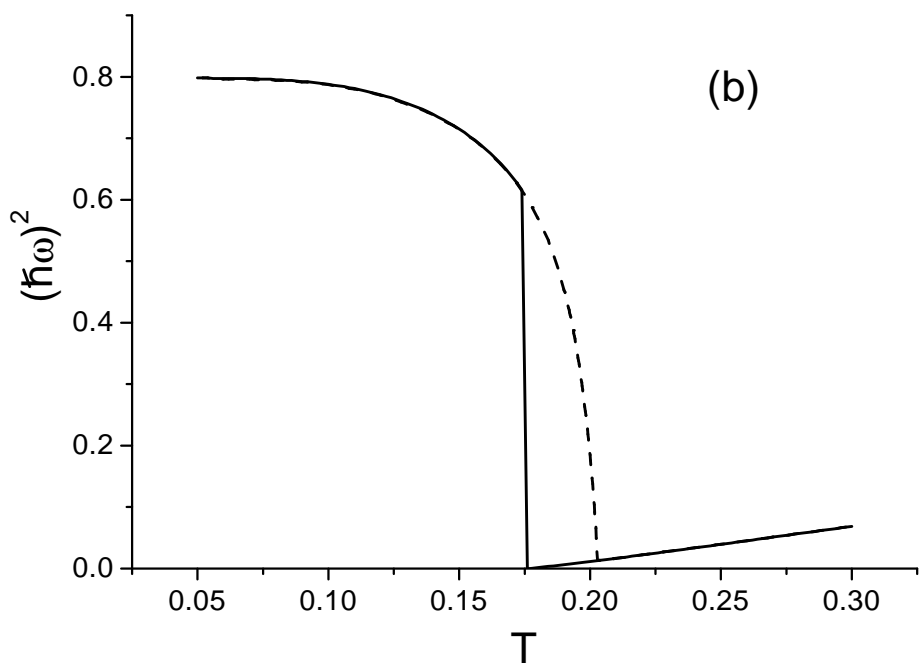
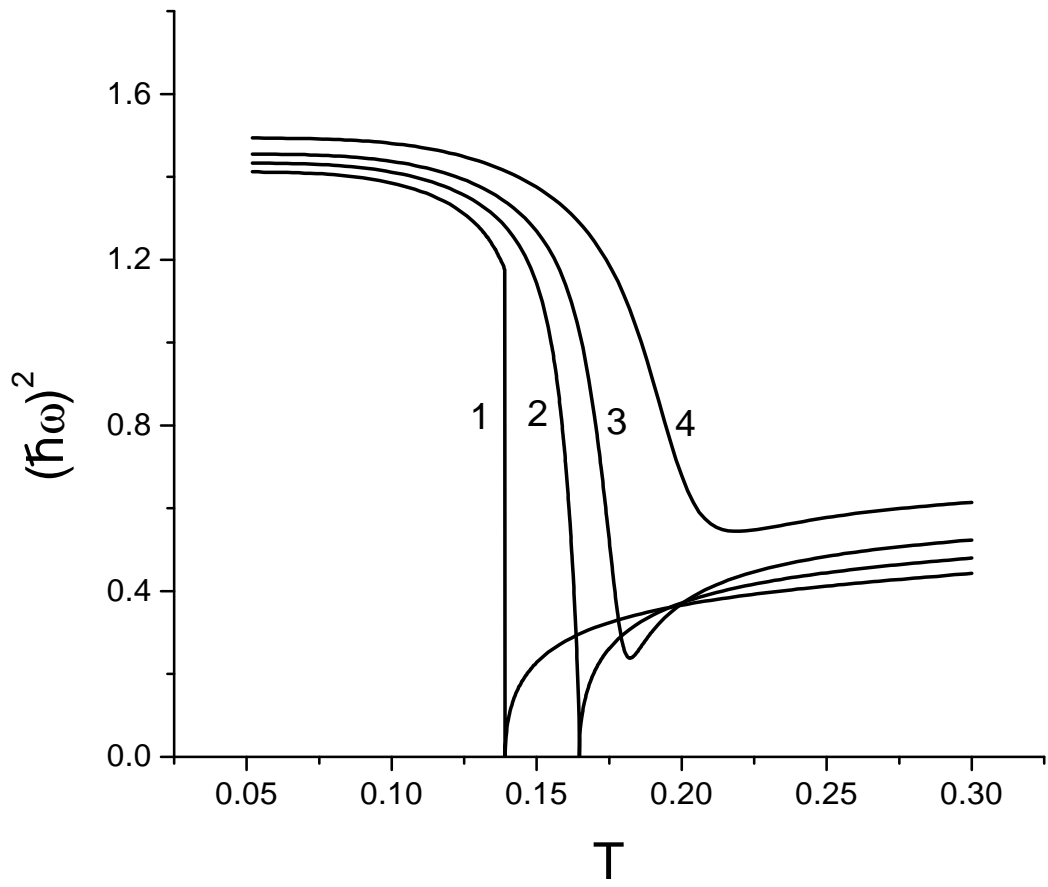
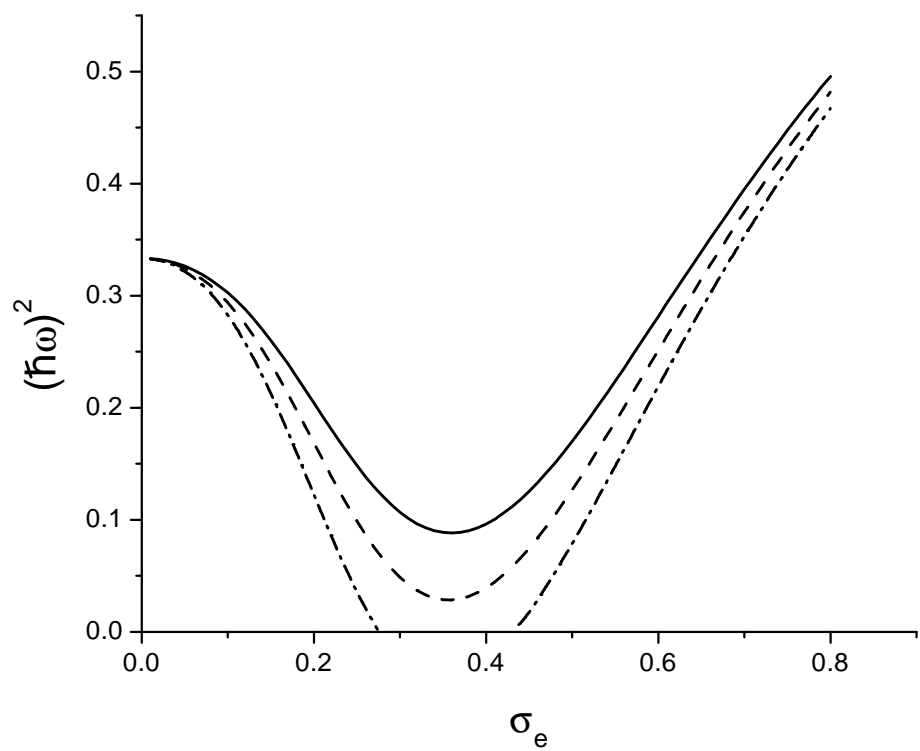
\end{document}